\documentclass[]{spie}  
\usepackage[]{graphicx}
\usepackage{url}

\title{TIPTOP: A NEW TOOL TO EFFICIENTLY PREDICT YOUR FAVORITE AO PSF} 

\author{Benoit Neichel\supit{a}, Olivier Beltramo-Martin\supit{a}, C{\'e}dric Plantet\supit{b}, Fabio Rossi\supit{b}, Guido Agapito\supit{b}, Thierry Fusco\supit{c,a}, Elena Carolo\supit{d}, Giulia Carl{\`a}\supit{b}, Michele Cirasuolo\supit{d}, Remco van der Burg\supit{e}
\skiplinehalf
\supit{a}Aix Marseille Univ, CNRS, CNES, LAM, Marseille, France; \\
\supit{b}INAF - Osservatorio Astrofisico di Arcetri Largo E. Fermi 5, 50125 Firenze Italy;\\
\supit{c}ONERA B.P. 72, F-92322 Châtillon, France;\\
\supit{d}INAF - Osservatorio Astrofisico di Padova Vicolo dell'Osservatorio 5, 35122 Padova Italy
\supit{e}European Southern Observatory, Karl-Schwarzschild-str-2, 85748 Garching, Germany 
}

\authorinfo{Further author information: Send correspondence to benoit.neichel@lam.fr}

\begin{document} 
  \maketitle 

\begin{abstract}
The Adaptive Optics (AO) performance significantly depends on the available Natural Guide Stars (NGSs) and a wide range of atmospheric conditions (seeing, Cn2, windspeed, …). In order to be able to easily predict the AO performance, we have developed a fast algorithm - called TIPTOP - producing the expected AO Point Spread Function (PSF) for any of the existing AO observing modes (SCAO, LTAO, MCAO, GLAO), and any atmospheric conditions. This TIPTOP tool takes its roots in an analytical approach, where the simulations are done in the Fourier domain. This allows to reach a very fast computation time (few seconds per PSF), and efficiently explore the wide parameter space. TIPTOP has been developed in Python, taking advantage of previous work developed in different languages, and unifying them in a single framework. The TIPTOP app is available on GitHub at: \url{https://github.com/FabioRossiArcetri/TIPTOP}, and will serve as one of the bricks for the ELT Exposure Time Calculator.

\end{abstract}



\section{INTRODUCTION}
\label{sec:intro}  
\subsection{Adaptive Optics Everywhere}
Adaptive Optics (AO) aims at compensating the quickly varying aberrations induced by the Earth's atmosphere. It overcomes the natural ``seeing'' frontier: the blurring of images imposed by atmospheric turbulence limiting the angular resolution of ground-based telescopes to that achievable by a 10 to 50cm telescope. Over the 20 past-years, AO for astronomy went from a demonstration phase, to a well-proven and operational technique. Today, all 8/10m telescopes are equipped with AO, and progressively turned into adaptive telescopes. The next step forward will come from the so-called Extremely Large Telescopes (39m diameter for the ELT\cite{ELT}, 30m for the TMT\cite{TMT}, 24m for the GMT\cite{GMT}) that will see first light in less than a decade. The scientific potential of these giants fully relies on complex AO systems, often integrated inside the telescope itself, and providing high-resolution images to all the instrumentation downstream. One crucial aspect for the science observations assisted by AO is the knowledge of the Point Spread Function (PSF). The PSF delivered by AO systems has a complex shape, combining spatial, spectral and temporal variability, such that it is difficult to predict. The AO PSF also highly depends on the atmospheric parameters and the Natural Guide Stars (NGSs) selected. Finally, the AO-PSF can also have a very different behavior depending on the AO flavor. The goal of this paper is to present a simple tool - called TIPTOP - which aims at simulating the expected AO PSF for any sort of AO system. In particular, this tool will be used in the frame of the ELT observation preparation, where the users will need to know the AO performance in order to properly design their observation strategies. 

\subsection{The ESO Working Groups}


ESO has initiated a series of Working Groups (WGs) to address the general problematic of “preparing observations with the ELT”. A detailed description of the WG can be found on the new ELT website: \url{https://elt.eso.org/about/}.  \\

The goal of these WGs is to provide the necessary infrastructure to prepare and execute observations with the ELT. Under this thematic, 4 WGs are respectively addressing the issue of:
\begin{itemize}
    \item Providing relevant star catalogs, down to H=21 which may represent the faintest stars to be used by AO instruments
    \item Providing tools to select the optimal asterism of NGSs \& Predicting the Adaptive Optics (AO) PSF based on those selected stars and the atmospheric conditions
    \item  Provide realistic numerical models of the instruments
    \item and eventually Providing an Exposure Time Calculator
\end{itemize}

Those four WGs are (logically) chained, each one getting information from the previous one, and providing inputs to the next one.\\
In this work, we focus on defining an algorithm capable of choosing the best combination of star(s) available in the field of view, and generate the PSF expected for the observation for a given AO system, a given observing mode and a given set of environmental conditions. In the case of the ELT, the exact requirements for this tool are defined in Section \ref{sec:TLRs}, but before going in these details, let us first define what the AO PSF is.

\subsection{The AO PSF}
 The 2D (x,y) AO-PSF formed at the focal plane of the scientific instrument is a function of wavelength ($\lambda$), time (t) and field position (r). To first order, it can be described by the convolution of three contributors: the telescope, the AO and the science instrument:

\begin{equation}
    \mathrm{PSF}(x,y,\lambda,t,r) = \mathrm{PSF}_{Telescope} * \mathrm{PSF}_{Atmosphere/OA} * \mathrm{PSF}_{Instrument}
\end{equation}

\begin{figure}[h!]
    \centering
    \includegraphics[scale=0.45]{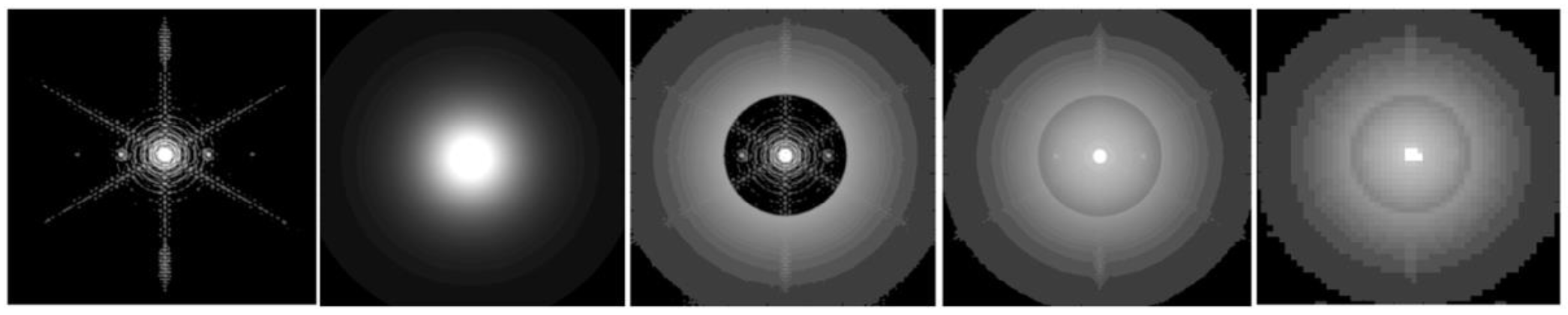}
    \caption{From left to right: the telescope PSF for an aperture including spiders and segments; a seeing limited PSF, a perfect AO-corrected PSF, a standard AO-PSF, the final PSF including pixel sampling and instrument aberrations.}
    \label{fig:fig_psfs}
\end{figure}

\noindent
\textbf{Telescope PSF:} The first term includes all the telescope specificities, the first one being the diffraction pattern imposed by the telescope aperture. For circular apertures, this would be the well-known Airy pattern, with a FWHM equal to $\lambda$/D, where D is the telescope diameter. This first term also includes the effect of central obstruction, spiders and all the telescope aberrations that will not (or partially) be corrected by the AO system, as for instance vibrations, windshake, field aberrations or phasing errors in case of segmented mirrors. Those aberrations are field, time and wavelength dependent and may affect all the PSF focal positions (i.e. all x,y – see Figure \ref{fig:fig_psfs}, left inset). As a first approach, these effects are encoded in TIPTOP via a phase mask. This phase mask includes the telescope pupil and a static shape to encode the telescope phasing issues. Vibrations and windshake can be added as a convolutional kernel. Some standard aberrations are proposed by default within TIPTOP, and users can also load their own if needed.\\

\noindent
\textbf{Atmosphere/AO PSF:} The second term depends on the atmospheric and AO system
characteristics. To first order, the AO system can be seen as a transfer function filtering the atmospheric perturbations. If there were no atmosphere, this PSF contributor would become a Dirac, and the resulting PSF would be independent of the AO system characteristics. This is the case for space missions, where the final PSF only depends on the telescope and instrument aberrations. At the other end of the range of the limiting cases, if the AO system is turned-off, this PSF becomes seeing-limited with a FWHM equal to $\lambda$/r0, where r0 (Fried parameter) encodes the atmospheric turbulence strength. Typical values of r0 are on the order of tens of centimeter, therefore, this atmospheric PSF is fully dominating the final PSF shape when compared to the telescope PSF (Figure \ref{fig:fig_psfs} – 2nd inset). The seeing-limited PSF is strongly time dependent, with variations faster than seconds, and with FWHM variations spanning ~0.3 to 2arcsecs for typical astronomical sites.

The AO system partially compensates for the aberrations induced by the atmosphere and the telescope. It is first important to understand that, because of the limited number of actuators on the AO deformable mirror, only a limited number of spatial frequencies can be corrected by the AO system. For instance, if the AO system were perfectly correcting all the aberrations within the range of its deformable mirror, the final PSF would be the combination of the Airy function near the optical axis and remains of the extended seeing-limited wings for focal positions above the correction range (Figure \ref{fig:fig_psfs} - middle inset). In reality, the AO system is not perfect and suffers from measurement noise, temporal or aliasing errors among others. Those error terms impact the PSF shape within the correction range and strongly depend on wavelength, field position and time (Figure \ref{fig:fig_psfs}, 4th inset). This is where TIPTOP will implement different filters for different AO systems. Some standard configurations corresponding to the different ELT instruments (see Section \ref{sec:elt_instrum}) are proposed, and the users will also have the possibility to play with the AO system parameters if they want to.\\

\noindent
\textbf{Instrument PSF:} This last term includes all the instrument characteristics, the first one being the sampling of the PSF by the detector pixels (Figure \ref{fig:fig_psfs} – right inset). But the scientific instruments may also carry their own aberrations, called NCPA (for Non-Common Path Aberrations). As for the telescope aberrations, part of those NCPA can be compensated by the AO system, and if these aberrations are static, they can be calibrated during day-time. A particular case applies to Integral Field Spectrographs, which can produce differential aberrations over the wavelength range, and for which the NCPA compensation can only be performed for a specific wavelength. Within TIPTOP, the users have the possibility to change the PSF sampling, or add static NCPAs maps if required.

\subsection{Adaptive Optics for the ELT}
\label{sec:elt_instrum}
Different spatial angular performance, hence different archetypes of AO-PSFs, are reached with different implementation of the AO modules. In the case of the ELT, all the instruments will implement AO. 
Indeed, one specificity of the ELT is to include a deformable mirror in its optical train: the fourth mirror (a.k.a. M4\cite{Vernet12}). This mirror has almost 6000 actuators that can be controlled at high temporal speed (up to 1000Hz), and all the instruments will make use of it. Depending on the science cases addressed by each ELT instruments, each is implementing a different AO flavor.

\begin{figure}[h!]
    \centering
    \includegraphics[scale=0.45]{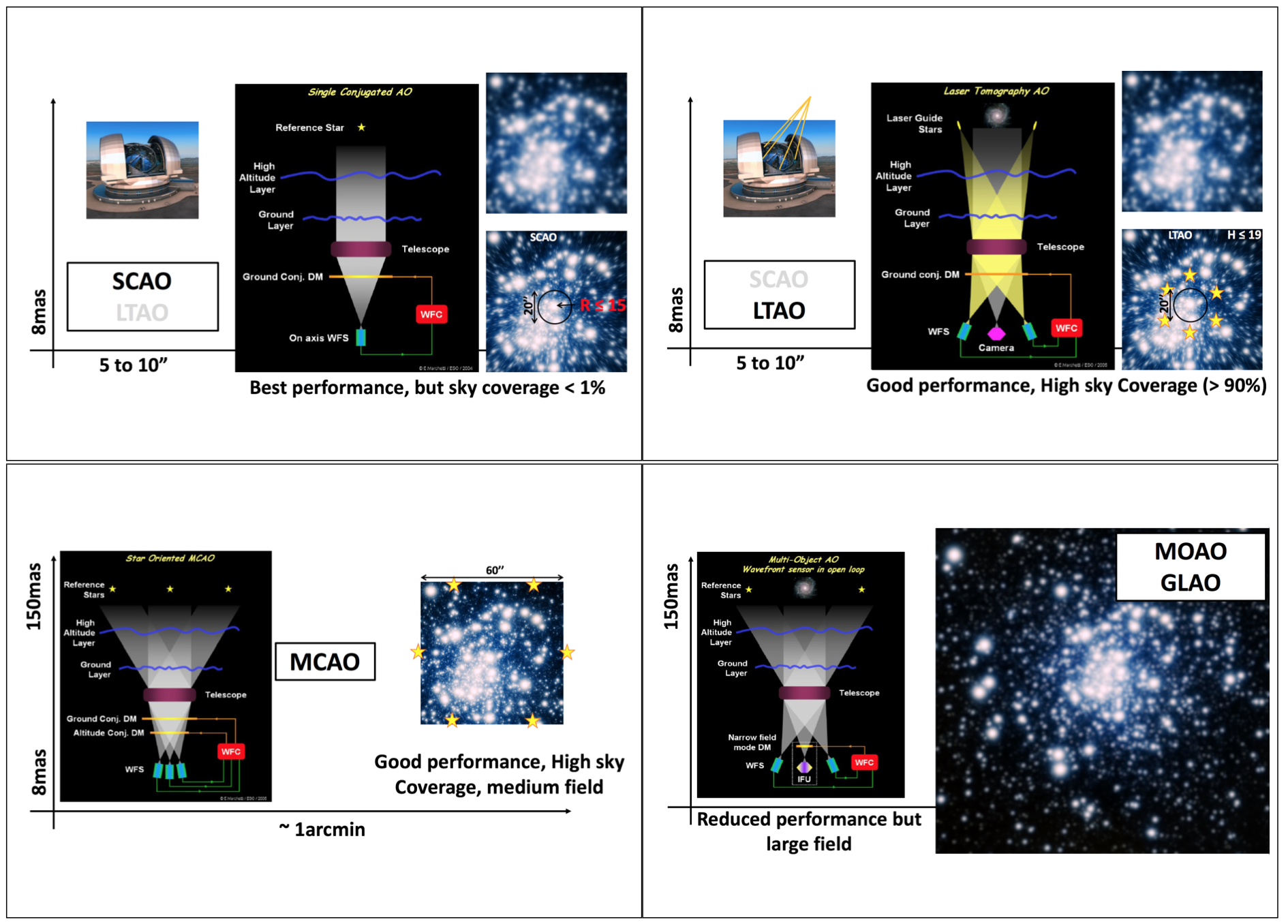}
    \caption{Illustration of the different AO flavors to be implemented for the ELT. {\bf Top-Left:} SCAO will be implemented by HARMONI, MOSAIC, MICADO and HIRES. {\bf Top-Right:} LTAO will be implemented by HARMONI {\bf Bottom-Left:} MCAO will be implemented by MAORY to feed MICADO. {\bf Bottom-Right:} GLAO and MOAO will be implemented by MOSAIC.}
    \label{fig:fig_aomodes}
\end{figure}

More specifically, HARMONI, MICADO and HIRES will implement Single Conjugate Adaptive Optics (SCAO - see figure \ref{fig:fig_aomodes} top left) systems \cite{Clenet18, Xompero18, Neichel16}. SCAO provides the best performance, brings the images to the diffraction-limit of the 39m telescope, but requires bright and close enough reference stars. Typically, a NGS with a magnitude brighter than R=14, and within a radius of $\sim$15arcsec should be used. For SCAO the shape of the PSF will mostly depend on integrated atmospheric parameters like the seeing, or the overall wind speed, and of course on the magnitude and off-axis distance of the NGS. The PSF shape also depends on the nature of the Wave-Front Sensors (WFS) ; all the SCAO systems planned for the ELT are using Pyramid WFSs. This is included within TIPTOP, and each instrument will have a specific configuration file to integrate its own specificity.\\

In order to tackle the sky-coverage issue of SCAO, the ELT will implement 6 Laser Guide Stars (LGSs) to allow for Laser Tomography AO (LTAO) (see figure \ref{fig:fig_aomodes} top right). An LTAO system provides almost similar performance as a SCAO system, however, over a fraction of the sky which is now almost complete. The sky coverage is not 100\%, because at least one NGS is still required to compensate for image motion at least. But this NGS may be fainter (typically H$<$19), and could be picked at a larger distance from the scientific target (typically 1 arcmin). HARMONI will implement an LTAO system \cite{Neichel16}. In this case, the shape of the PSF depends on the vertical structure of the turbulence, including seeing and wind speed, and also the position and magnitude of the NGS. This is all encoded into TIPTOP. \\

An LTAO system solves the sky-coverage limitation, but the correction provided is only optimized for a small Field of View (FoV -  typically less than 10 arcsec). To increase the corrected FoV, it is necessary to implement post-focal deformable mirrors, that are used in conjunction with M4. With more deformable mirrors, and the several LGSs, an MCAO system can deliver diffraction limited performance over a field that can reach 1 or 2 arcminutes (see figure \ref{fig:fig_aomodes} bottom left). MAORY is the MCAO module of the ELT \cite{10.1117/12.2234585,ciliegi2020maory}. It will feed MICADO. The PSF shape also depends on the vertical structure of the atmosphere, and the magnitude / location of the required NGSs.\\

If one wants to significantly increase the corrected FoV, trade-offs are to be made on the level of correction provided by the AO system. With a single deformable mirror, as is M4, but combining WFSs measurements from far off-axis LGSs or NGSs, the system is called Ground Layer AO (GLAO). The level of correction provided by a GLAO system will be partial, but uniform over a large field (see figure \ref{fig:fig_aomodes} bottom right). A GLAO system only compensates for the atmospheric turbulence in the first hundreds of meters above the telescope, but those are usually the most energetic ones. As such, a GLAO correction will not provide diffraction limited images, but typically shrink the seeing PSF image by a factor 2 to 5. It can be seen as seeing-reducer, shifting the median seeing of Armazones from $\sim$0.65 arcsec down to $\sim$0.2 or 0.3arcsec. MOSAIC intends to use a GLAO correction for its High-Multiplex Mode (HMM\cite{Morris18}). In this case, the shape of the PSF mostly depends on the fraction of the turbulence near the ground\cite{Fusco20}.\\

Finally, one way to improve the performance over a very large FoV is to provide local corrections, with dedicated deformable mirrors. This is called Multi-Object AO (MOAO). MOAO systems are mostly driven by extra-galactic science cases, where it is not needed to have a full corrected FoV, but only focus on specific directions: where the galaxies are. MOSAIC intends to implement an MOAO correction for its High-Definition Mode (HDM\cite{Morris18}). \\

The goal of TIPTOP is to provide the estimated AO-PSFs for all these AO configurations, in a fast enough way so that users can predict the performance for as many configurations as needed. The exact requirements for TIPTOP are described in the next section.

\section{Top Level Requirements for an AO-Prediction tool} 
\label{sec:TLRs}
Mostly motivated by the needs of the ESO WG, we have defined what the TIPTOP tool shall (or not) deliver. This is summarized below.\\
First, we recall that the deliverables of TIPTOP are twofold: 
\begin{enumerate}
    \item Starting from a given catalog of stars, including star positions and magnitude, TIPTOP shall rank the possible asterisms by their expected performance.
    \item Based on a given set of system, atmospheric and NGSs parameters (set to be defined), TIPTOP shall provide the expected AO-PSF.
\end{enumerate}

The AO-PSFs are provided over a grid, and at the wavelengths provided by the users. The PSF spatial and spectral sampling is also a free parameter adjusted by the user. By default, we should be able to provide any sampling, wavelength, field position, if we assume the right inputs are provided. By default, the PSFs produced are long (infinite) exposure PSFs. It will be possible to also generate short exposure PSFs, but this will come as a second step.

One important requirement for TIPTOP is to be able to generate PSFs quickly. Typically, the goal is to have a tool that can generate AO-PSFs on-the-fly, with an output produced in less than a few seconds. This directly impacts the choice of the algorithms, and the final implementation, as explained in Section \ref{sec_python}.

In terms of final performance, it is important to note that the main motivation for this work is to provide AO-PSFs for an Exposure Time Calculator (ETCs), and not to do science analysis. Hence if the final AO-PSF accuracy is in the order of few percent, it should be considered as a satisfactory result. The exact requirement on accuracy is still under construction, and for that some detailed simulations are carried out to understand the impact of the PSF shape on the final astrophysical SNR. Of course this depends on each and every science case, so this task will be an on-going work in the following years. Typical performance achieved with the current version is described in Section \ref{sec_perfo}.

If the TIPTOP algorithm is fast enough, then it will be possible to provide a range of PSFs around the observational inputs, spanning the multi dimensional parameter space of environment conditions (r$_0$, Tau$_0$,Theta$_0$, L$_0$, Sodium content, etc...). This may be used to provide error bars on the estimated PSFs and/or assess the feasibility of an Observing Block.

Finally, if the tool can be fast enough, it may be used during the night for queue planning. Based on the current atmospheric conditions, or based on the predicted conditions of a weather forecast algorithm, the night-time operator may be able to predict the associated AO performance, and select the best instrument accordingly.

\section{Basic Strategy} \label{sec:strategy}
There are several ways of computing AO PSFs, but as the main requirement was to be fast, we focused our strategy around analytical tools, computing the PSF from a residual phase Power Spectral Density (PSD) in the Fourier domain. For that, we recycled the work from Neichel et al. 2008\cite{Neichel2008} , and Plantet et al. 2018\cite{Plantet18} .

\begin{figure}[h!]
    \centering
    \includegraphics[scale=0.7]{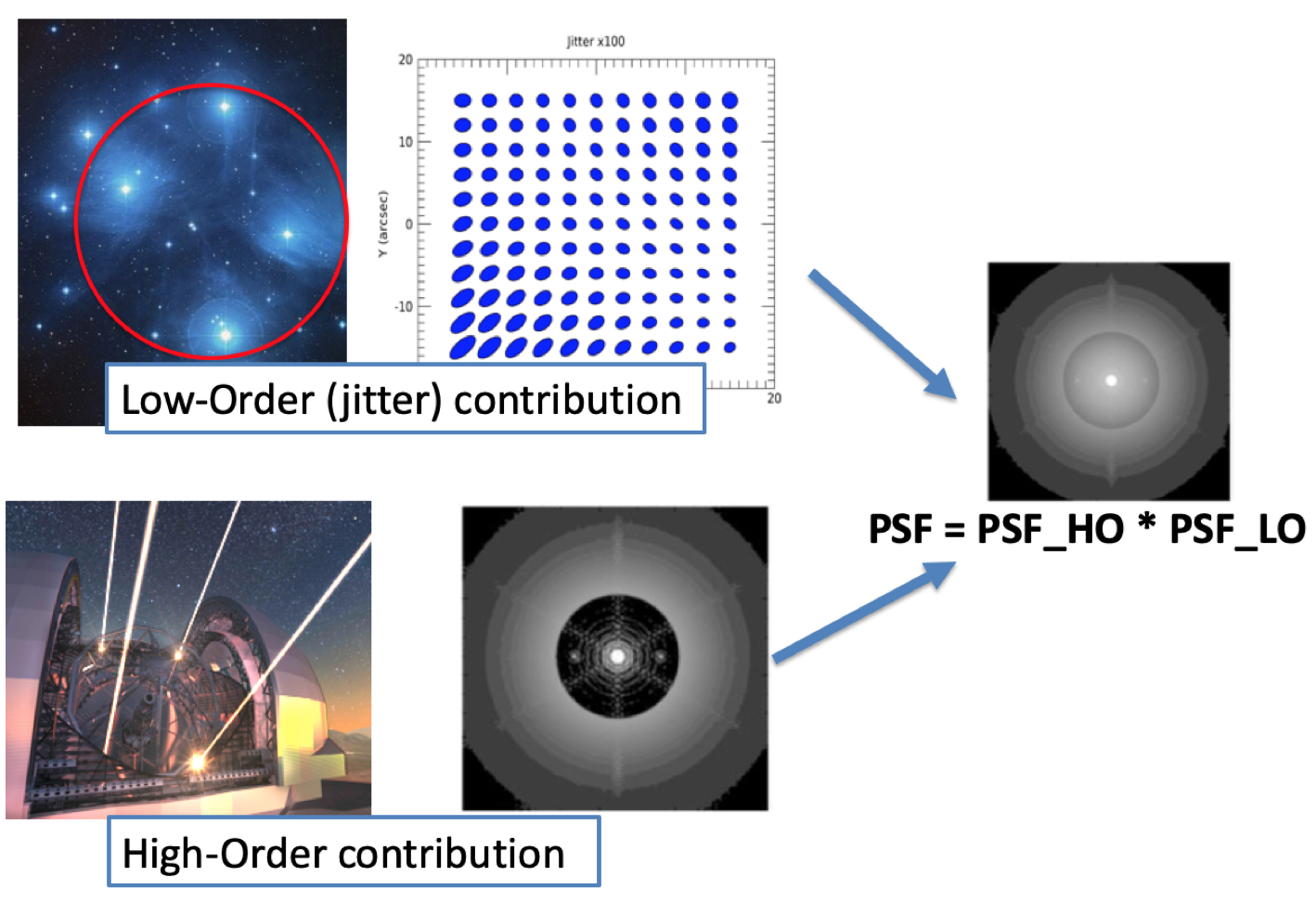}
    \caption{Schematic description of the TIPTOP strategy. One one hand, and based on a star catalog, the expected jitter is computed. This output can be used to test different NGSs asterism, and select the best one, based on user criteria (e.g. more uniform jitter, best peak performance, etc...). In parallel, the high-order part of the PSF is computed, which is mostly fixed by the system and atmospheric inputs. The two parts are eventually convolved to form the final PSFs, which can be estimated over the field, and at different wavelengths.}
    \label{fig:fig_strategy}
\end{figure}

The strategy is to decouple the High-Order (HO) part of the PSF, which only depends on the LGS constellation and atmospheric conditions, from the Low-Order (LO) part (a.k.a. the jitter), which strongly depends on the chosen NGS asterism. From a schematic point of view, this strategy is described by Figure \ref{fig:fig_strategy}. Both parts are computed in parallel:
the HO produces a PSF including all the tomography/telescope aspects. The LO produces a map of jitter across the field, for each of the NGS asterism combinations. The user can then select the best NGSs for his observation. The final PSF is then produced by convolving the HO PSF with the jitter kernel.

\begin{figure}[h!]
    \centering
    \includegraphics[scale=0.5]{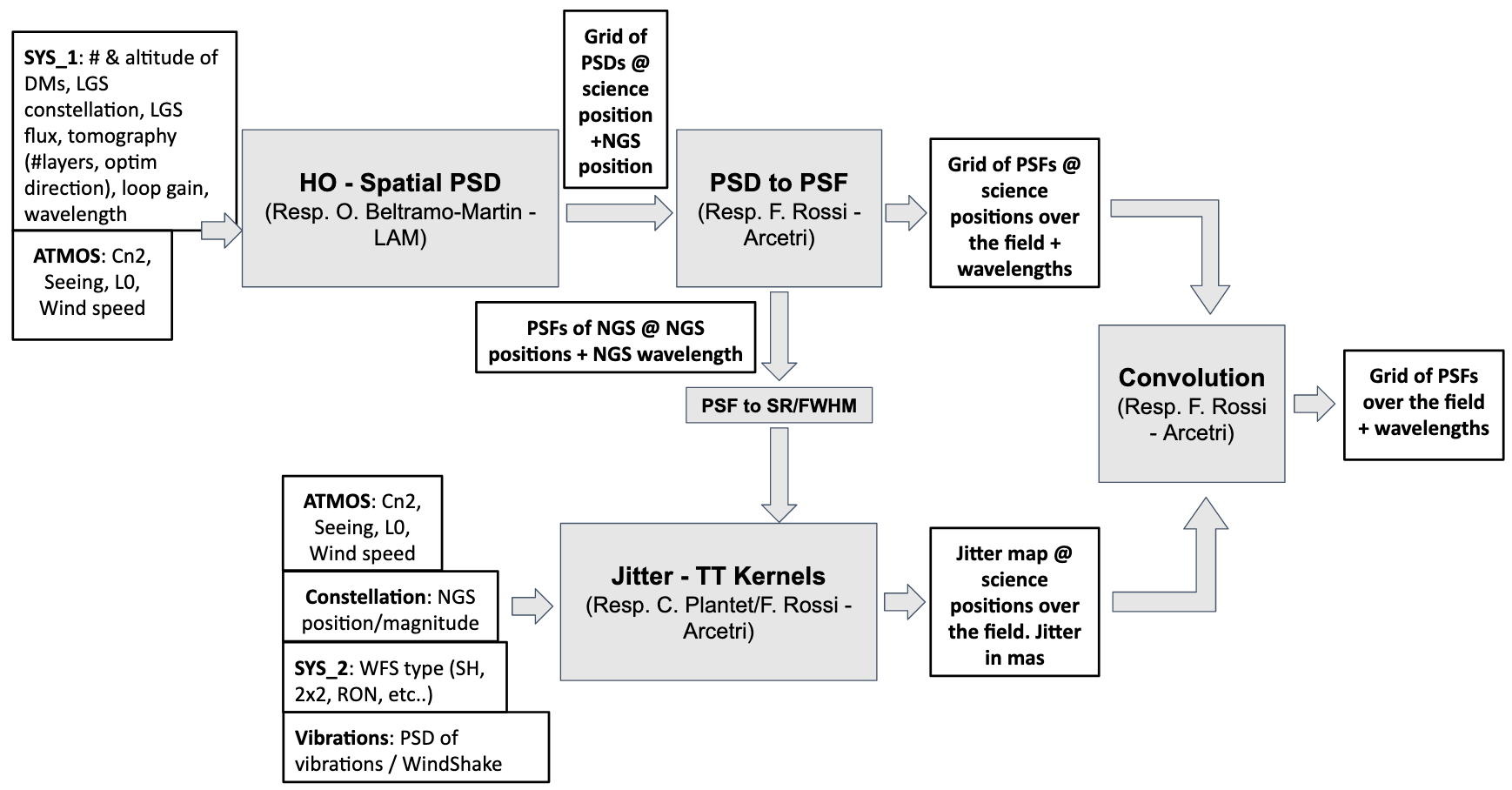}
    \caption{Block diagram schematic of the different steps computed within TIPTOP. White boxes are the inputs, and grey boxes are the algorithm bricks. Each part can be called independently, or the full sequence can be ran at once.}
    \label{fig:fig_algo}
\end{figure}

In fact, the path is slightly more complex, and is described by Figure \ref{fig:fig_algo}. The high-order part computes PSDs of the residual phase, for each field direction, but it also computes it for the specific NGS directions. This is required as one of the inputs for the Low-Order computation is the residual phase variance (in fact the PSF shape with its Strehl Ratio SR and FWHM) in the NGSs directions. This grid of HO PSFs can be computed at several wavelengths, and any directions. 

\section{High-Order part of the PSF}
The high-order part of the PSF is computed from the PSD of the AO-corrected phase $\mathrm{PSD}_{Atmosphere/AO}$  following the scheme presented Neichel et al. 2008\cite{Neichel2008}:
\begin{equation}
    \mathrm{PSF}_{Atmosphere/AO} \propto \mathcal{F}^{-1}(\exp(\mathcal{F}\left[\mathrm{PSD}_{Atmosphere/AO}(\mathbf{k})\right])),
\end{equation}
where $\mathcal{F}(\mathbf{x})$ is the 2D Fourier transform of $\mathbf{x}$ and $\mathbf{k}$ the spatial frequencies domain, and $\mathrm{PSD}_{Atmosphere/AO}$ is derived from a combination of PSDs of specific AO errors that are assumed to be independent:
\begin{equation}
    \mathrm{PSD}_{Atmosphere/AO} = \mathrm{PSD}_{Fitting} + \mathrm{PSD}_{Noise}+ \mathrm{PSD}_{Aliasing} +\mathrm{PSD}_{Spatio-temporal},
\end{equation}
where each contribution is detailed below:
\begin{itemize}
\item \textbf{Fitting}: refers to the uncorrected high-spatial frequencies above the AO correction radius, e.g. $1/(2\times pitch_{act})$ in the PSF domain, where $pitch_{act}$ is the DM actuators pitch in meters. The transition is generally assumed to be circular to account for filtered modes; the user can setup a squared transition if desired, though. This term only depends on the seeing value (this term is poorly L0-sensitive) at the reference wavelength, which are user-defined parameters.
\item \textbf{Noise}: refers to the WFS noise (detector, shot noise, background) that creates a signal that propagates through the AO loop and affects the PSF. The model accounts for the (tomographic) wavefront reconstruction and the AO loop temporal model. The use can either provide WFS characteristics (pixels/sub-aperture, read-out-noise, total throughput, pixel scale) that will feed a noise variance calculator following \cite{Rousset1987} formulas, or directly provide the noise variance in $rd^2$.
\item \textbf{Aliasing}: refers to the high-spatial frequencies that are aliased owing to the WFS spatial sampling and that propagate through the AO loop as well. The calculation of this term is particularly demanding for tomographic systems, and in order to speed it up, the code systematically computes the aliasing in a SCAO scenario and does not account for the propagation through the tomographic reconstructor and the projector. In practice, the PSF shape is weakly sensitive to this approximation.
\item \textbf{Spatio-temporal}: refers to the spatial error (wavefront reconstruction, tomography, DM projections, anisoplanatism for SCAO systems) that is combined with the temporal error (loop bandwidth, delays) into a single term. The exact calculation of this term is complex and given in Ref. \citeonline{Neichel2008}. The user must define the positions and altitude (for LGSs) of guide stars, as well as the altitude conjugations/actuators pitch of the DM and the optimization directions. The tomographic error is calculated in the context of pseudo-open-loop command (POLC) and Minimum Mean Square Reconstruction (MMSE) only. If LGSs are considered, the atmospheric layers are stretched to account for the cone effect. 
\end{itemize}

\section{Low Order part of the PSF}
The approach to compute the low-order residuals is mostly based on the method presented in Plantet et al. 2018 \cite{Plantet18}, which was designed for MAORY. The residual jitter is considered as the quadratic sum of 3 independent terms:
\begin{itemize}
    \item \textbf{Windshake/vibrations}: The wind on the telescope and/or the instrumentation itself can produce strong vibrations, that will mostly be seen as a jitter. The correction of these vibrations needs to be fast, as they might have significant power at high frequencies. On the other hand, since everything happens at the level of the telescope, this jitter is isoplanatic. We can thus consider a SCAO-like case on the brightest NGS to compute this error term. The residual is derived from the expected temporal PSD of the vibrations (for now purely theoretical) on which we apply a temporal filter corresponding to an AO control law, e. g. a double integrator. The temporal filter parameters (gain, loop frequency...) are optimized with respect to the SNR expected on the brightest NGS.
    \item \textbf{Tomographic error}: This error is due to the difference between the turbulence volume on the line of sight of the science camera and the ones on the lines of sight of the NGSs. It only depends on the NGS asterism geometry. The contribution of the tomographic error is computed from the formulas given in Plantet et al. 2018 \cite{Plantet18}. The formulas can easily be adapted to any direction in the scientific FoV and/or to the case of a single NGS, for which the residual jitter becomes the classical anisoplanatism error.
    \item \textbf{Noise error}: The noise error corresponds to the propagation of the NGS sensors' noise (photon noise, detector noise\dots) through the LO loop. The jitter error is analytically computed for each sensor in the same way as it is done for the windshake in Plantet et al. 2018 \cite{Plantet18} (see section 2.1 and appendix B): we compute the slope error from a simple Gaussian model of the PSF, and then propagate the noise through a simple integrator loop. If there is only one NGS, then the noise error is directly the result of that computation. If several NGSs are used, then we also need to propagate the error through the tomographic reconstructor. This latter propagation is detailed in appendix C of Plantet et al. 2018 \cite{Plantet18}, together with the tomographic error.
\end{itemize}

\section{Python implementation}
\label{sec_python}
We developed the simulation software in Python, making an effort in its design and implementation to provide the option to run its computationally intensive parts either on CPU or GPU (Nvidia CUDA enabled). We followed the approach described in Rossi 2020 \cite{rossi-adass-2020}: mathematical formulas were at first specified as SymPy expressions, allowing easy verification and preliminary checks on their correctness. Then, such symbolic expressions were translated automatically to the Array Programming backend of choice (NumPy for CPU or CuPy for GPU in our case) by the SEEING (Sympy Expressions Evaluation Implemented oN the GPU) library, to be finally used in the backend agnostic numerical code.
The final simulations were produced on a machine equipped with an NVIDIA TITAN X graphics card, providing a speedup between one and two orders of magnitude compared to execution on CPU only. All simulation parameters are configurable and stored in a .ini file.
The software is available on GitHub at: \url{https://github.com/FabioRossiArcetri/TIPTOP}.

\section{TIPTOP Performance}
\label{sec_perfo}

\begin{figure}[h!]
    \centering
    \includegraphics[scale=0.7]{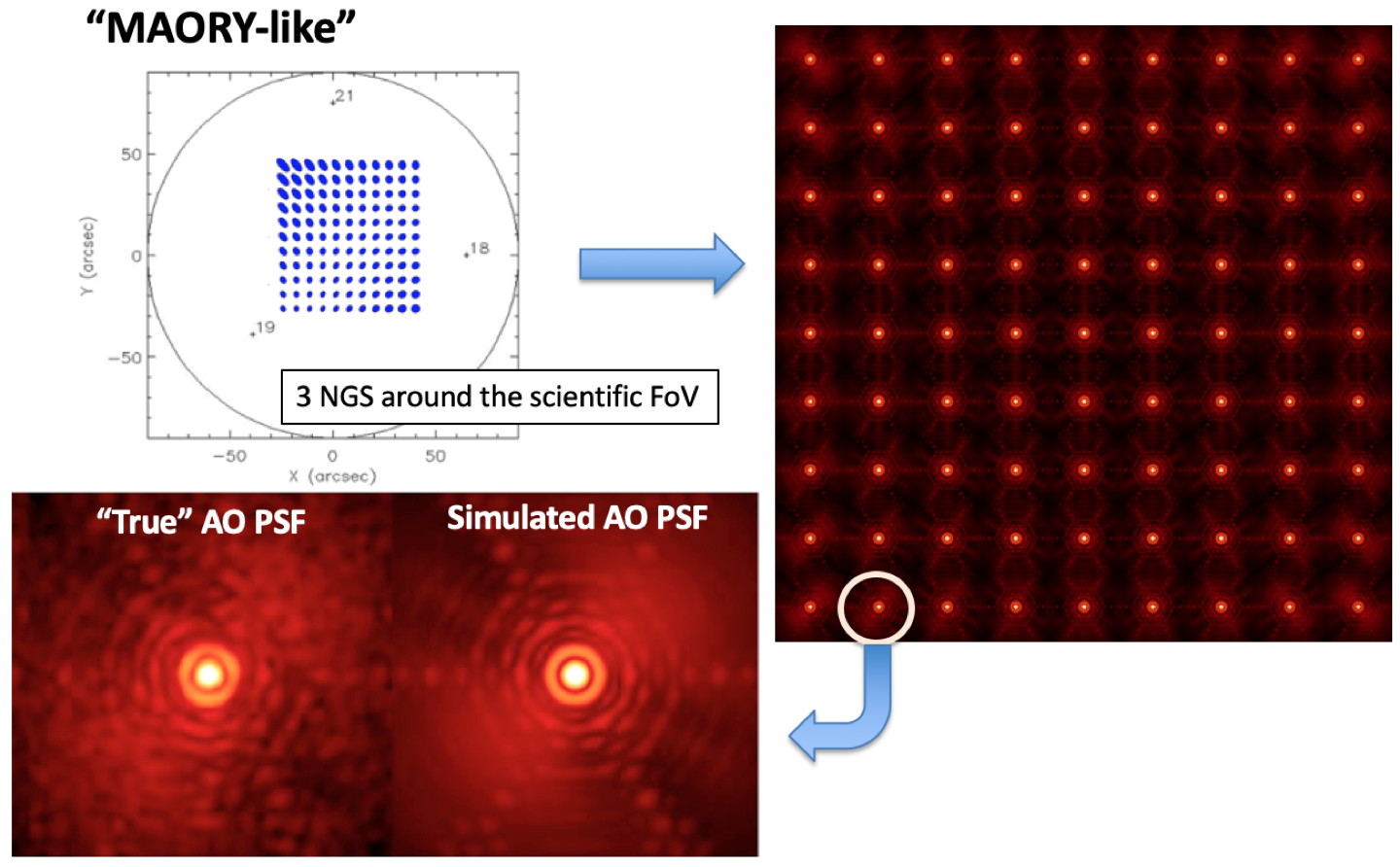}
    \caption{Illustration of an output produced by TIPTOP, as a grid of PSFs over a 1 arcmin FoV. In this specific case, we simulate a MAORY configuration, with 3NGSs of magnitude H=18, 19 and 21, and located around the scientific field. The lower inset shows the comparison between a PSF obtained with the PASSATA E2E code, considered as the ``true'' PSF here, and the TIPTOP output. It can be seen that the main features of the PSFs are properly reproduced, and the main differences come from the finite exposure time of the E2E code. The big advantage of TIPTOP, is that it only took few seconds to compute all the PSFs, while it would have taken several hours for an E2E tool.}
    \label{fig:fig_results1}
\end{figure}

We have started to test the accuracy of the outputs produced by TIPTOP. For that, we compared the PSFs produced by TIPTOP with PSFs obtained from End-to-End (E2E) codes, as for instance PASSATA\cite{doi:10.1117/12.2233963} or OOMAO\cite{10.1117/12.2054470}. A first result is shown in Figure \ref{fig:fig_results1}, in this case for a MAORY-like configuration. In this configuration, 3 NGSs of magnitude H=18, 19 and 21, located around the scientific field are considered. The associated jitter map then shows the typical expected elongation. Once convolved with the grid of HO-PSF, the final result is the grid of PSF shown on the right of Figure \ref{fig:fig_results1}. A zoom-in on the output, and a comparison with the PASSATA PSF (used as a reference ``true'' PSF) is also shown. In this example it can be seen that results are quite close, and that the overall PSF morphology is well modelized by TIPTOP. The main difference between the E2E and the analytical PSF mostly comes from the limited exposure time simulated with PASSATA, and the associated speckles. \\

\begin{figure}[h!]
    \centering
    \includegraphics[scale=0.7]{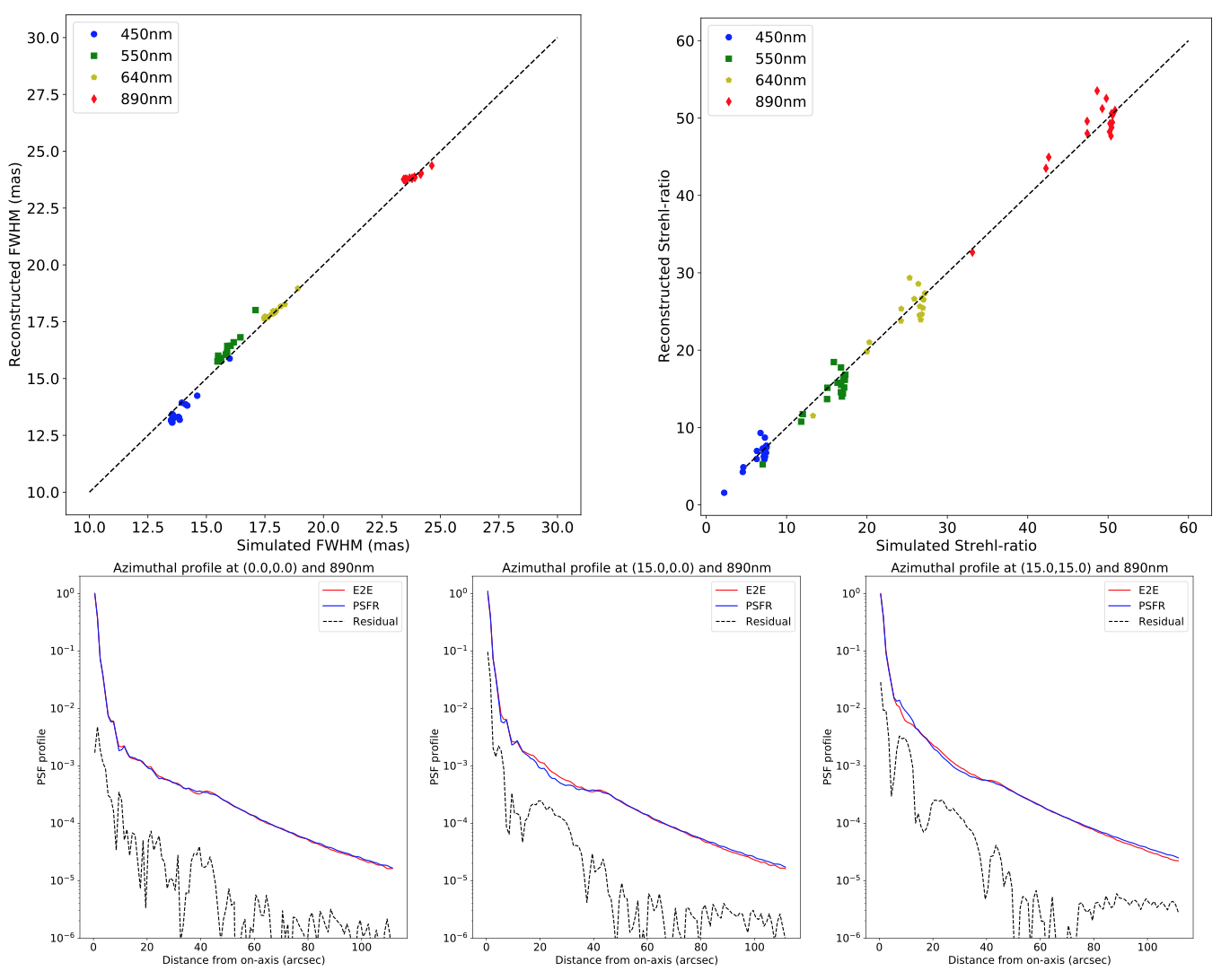}
    \caption{Comparison between PASSATA E2E and TIPTOP analytical outputs, for a MAVIS configuration. Top row shows the comparison for the FWHM (left) and SR (right), and the bottom row shows some radial cut of the PSFs over the field, respectively at the center of the science field (0,0), 15arcsec off axis along the X-direction (15,0) and 15arcsec off-axis along the 2 axis (15.0, 15.0). The overall agreement between the two tools is better than few percent, and fulfills the performance requirement.}
    \label{fig:fig_results2}
\end{figure}

\begin{figure}[h!]
    \centering
    \includegraphics[scale=0.7]{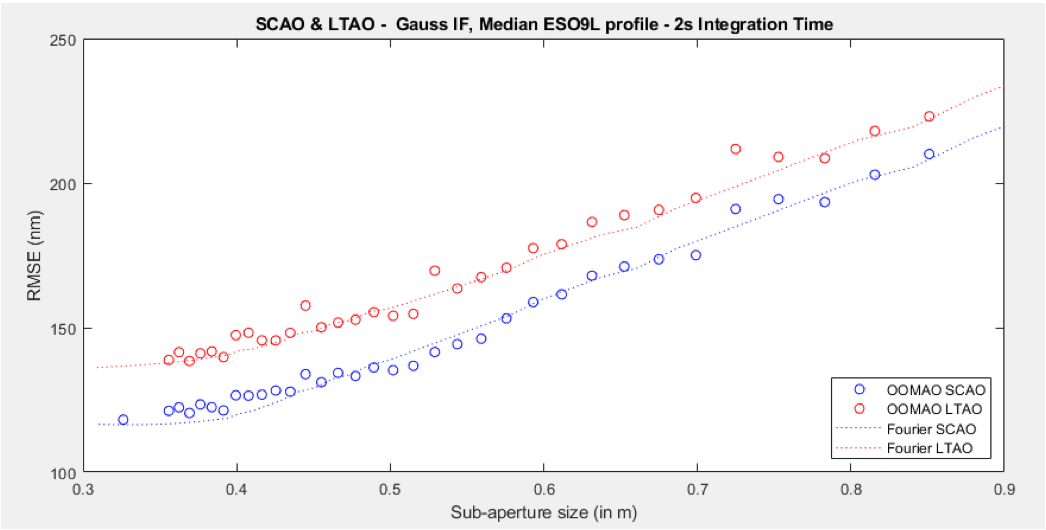}
    \caption{Comparison between OOMAO E2E and TIPTOP outputs for an HARMONI SCAO and LTAO configuration. The residual variance vs. WFS sub-aperture size is used as the metric in this case. The agreement between the 2 codes is again very good.}
    \label{fig:fig_results3}
\end{figure}

We have performed more comparison between E2E and TIPTOP PSFs, shown in Figure \ref{fig:fig_results2} and \ref{fig:fig_results3}. In Figure \ref{fig:fig_results2} we considered a MAVIS configuration, to speed up the E2E simulations. MAVIS is a 3rd generation MCAO instrument for the VLT, working in the visible. It will use 8 LGSs, 3 DMS and 3NGSs. Results of Figure \ref{fig:fig_results2} show the relationship between two PSFs metrics - the FWHM and the SR - when computed for an E2E PSFs and from TIPTOP. This is done at different wavelengths. The agreement is quite good, with less than a few percents of error. The bottom row of Figure \ref{fig:fig_results2} shows some examples of PSF radial cut, along with the residual. Again the agreement between the E2E results and the fast analytical TIPTOP is quite good, with less than a few percents of error. These first tests are very  encouraging and validate our approach.

Finally, a last example of the good behavior of the TIPTOP output is shown in Figure \ref{fig:fig_results3} for an HARMONI SCAO and LTAO configuration. In this case, we compared the phase variance (strongly related to the SR) produced by the OOMAO E2E tool and the TIPTOP output, depending on the WFS sub-aperture size. Again, the agreement with this other E2E tool is quite impressive, and validates the analytical approach.

Finally, it is important to note that the TIPTOP tool can easily be tweaked to adjust the output performance, and once the first AO systems will be on-sky, it will be possible to calibrate the software to get the best match as possible with real observations.

\section{Conclusions}
TIPTOP is a new tool able to quickly, but efficiently simulate any AO-PSF. Its main purpose will be to be integrated in a pipeline of observation preparation for the ELT. As such, TIPTOP produces any kind of AO-PSFs (SCAO, LTAO, MCAO, GLAO, MOAO) in only a few seconds. These PSFs can be computed at any sampling, position in the field and wavelength. As such, it is a very flexible tool, which may be useful for other applications. The overall implementation has been done in Python, making use of GPU computing power, and the software is available on GitHub at: \url{https://github.com/FabioRossiArcetri/TIPTOP}.

\acknowledgments     
 
This work has been prepared as part of the activities
of the French National Research Agency (ANR) through the ANR APPLY (grant ANR-19-CE31-0011).
Authors also acknowledge the support of OPTICON H2020 (2017-2020) Work Package 1 (Calibration and test tools for AO assisted E-ELT instruments). OPTICON is supported by the Horizon 2020 Framework Programme of  the  European  Commission’s  (Grant  number  730890). Authors are also acknowledging the support by the Action Spécifique Haute Résolution Angulaire (ASHRA) of CNRS/INSU co-funded by CNES, and the support of LABEX FOCUS ANR-11-LABX-0013. 

\bibliography{report}   
\bibliographystyle{spiebib}   

\end{document}